\documentclass[aps,prl,twocolumn,groupedaddress]{revtex4}
\usepackage{amsfonts}
\usepackage{amsmath}
\usepackage{amssymb}
\usepackage[dvipdfmx]{graphicx}
\usepackage{braket}
\usepackage{amsmath}
\usepackage{amsfonts}
\usepackage{amssymb}
\usepackage{color}%

\allowdisplaybreaks[3]

\providecommand{\U}[1]{\protect\rule{.1in}{.1in}}
\providecommand{\U}[1]{\protect\rule{.1in}{.1in}}
\begin{document}
\title[Antisymmetrized Geminal Powers with Larger Chemical Basis Sets]
{\textbf{Antisymmetrized Geminal Powers with Larger Chemical Basis Sets}}
\author{Wataru Uemura and Takahito Nakajima}

\affiliation{RIKEN Center for Computational Science, Kobe 650-0047, Japan}
\pacs{PACS number}

\begin{abstract}
In previous research, we tested the wave function format of 
a linear combination of several antisymmetrized geminal power states.
A numerical problem 
in the geminal matrices was noted, which made the total energies of electronic systems with large numbers of electrons unstable.
The underlying cause was found to be the 
large cancellation term in the geminal power series.
We have obtained a new format to resolve this 
problem for the case of total energies and 
partly for the first-order derivatives
within the antisymmetrized geminal power states.
By using this new formalism, we have 
calculated the ground state energies for several
electronic systems, including the usage of 
a larger chemical basis set.
The results are, in some cases, very close to the 
exact result, especially for 
one-dimensional Hubbard systems. Our result for a water molecule with the Dunning Zeta basis set 
is better than the CISD energy and approaches the CCSD energy.

\end{abstract}

\maketitle

There are many fascinating physical phenomena
in macroscopic compounds
of transitional metallic elements like cuprate superconductors.
These physical phenomena are also chemical and
governed partly by quantum chemical mechanisms.
To obtain the energy
spectra of these chemical systems,
we have used the configuration interaction
method or the coupled cluster formalisms\cite{shavitt, saxe, purvis}.
These methods have polynomial 
scaling in computational cost 
versus system size, but attempts at obtaining systematically good results 
increased the cost index of the 
scalings.\par
There are many wave function theories 
designed to conquer the problems of 
past decades, including the density matrices 
renormalization group methods (DMRG)\cite{white1992}\cite{white1993}.
The algorithm of DMRG 
is one example
of matrix product states.
When there are 
one-dimensional structures inherent to the electronic 
system, DMRG provides very good results\cite{shollwock}.
When applying the DMRG format
to real molecules, 
we must sometimes assume a
one-dimensional order of the chemical 
sites of 
the systems, which becomes increasingly
unnatural when we increase the system size\cite{chan2002}.\par
For the case of electronic systems,
we have the knowledge of 
quantum Monte Carlo algorithms (QMC)\cite{foulkes}.
Some of these algorithms use Slater determinants, 
and in some cases are further extended to 
use the so-called pfaffian or the 
antisymmetrized geminal power states (AGP)
with Jastrow factors (JAGP).
The combination of geminal powers 
and Jastrow factors has already
obtained a variety of results for chemical systems\cite{bajdich2006, bajdich2008, eric, eric2, sorella}.
The formalism of QMC is based on
Walker sampling with the usage of probability;
thus, in obtaining the expectation value 
of the energies, we use the statistical average\cite{booth}\cite{haoshi}.\par
When using electronic functions larger than 
a one-electron density, we sometimes 
use the second-order reduced density matrices\cite{coleman1}\cite{coleman2}.
The energy formula of the density matrices is very
trivial in comparison to DFT formalisms,
but there are known problem such as the N-representability
of the density matrices\cite{verstraete}.
There are many 
ways to use the special conditions of the 
density matrices to conquer these problems\cite{nakata}.
There are also suggestions for usage of 
antisymmetrized geminal power states, which are 
an extension of Slater determinants 
and very close analogues to the well-known 
BCS states\cite{coleman1, coleman2, Ortiz1981, weiner, giorgini}.
These AGP states are used to obtain the potential surfaces
of molecules\cite{Straroverov2002}\cite{Scuseria2011}.\par
We have previously introduced the wave function 
assumption of the linear combination of 
antisymmetrized geminal power states
to correctly and compactly describe
the structures of many electronic wave functions\cite{uemura2012}\cite{uemura2015}.
We hoped this format would 
provide an established workplace 
to describe larger chemical and physical systems in the future
with a polynomial cost and high precision results.
When we use the linear combination of Slater determinants instead of AGPs
we can, in some cases, obtain results 
comparable to the exact value\cite{Goto}.
AGPs describe larger variational spaces 
than the Slater determinants, so 
they are likely to provide better results.
There are also recently reported results for geminal powers, 
and some provide results for the
geminal product case\cite{katharina}\cite{katharina2}, while others provide 
results for the restricted Hamiltonian case\cite{kawasaki}.\par
On the practice of the geminal power states with the 
electronic systems, we found that a numerical instability arises due to the 
matrices power structure of the format.
We found that the cause of the problem is the 
cancellation of high-power terms when we 
obtain the total energy formula.
We introduce how we obtained the solution in the 
forthcoming section.
Furthermore, we report on recent
calculation results with several antisymmetrized 
geminal power states.
The geminal matrices technique was used,
and we successfully enlarged the area of the 
system sizes such that we could calculate with the 
geminal power states.\par
First, we present the basic expressions in our AGP-based formalism. Some parts of these expressions 
have appeared in ref \cite{uemura2015}.
We give the expression of the wave function as 
\begin{equation}
\Psi\left(  x_{1}\cdots x_{N}\right)  =\sum_{i_{1}\cdots i_{N}=1}^{M}%
A_{i_{1}\cdots i_{N}}\phi_{i_{1}}\left(  x_{1}\right)  \cdots\phi
_{i_{N}}\left(  x_{N}\right),
\end{equation}
\begin{equation}
A _{i_1\cdots i_N}=\sum _{k=1}^{K}c_k
\hat{A}(\gamma_{i_{1}i_{2}}^{k}\gamma_{i_{3}i_{4}}^{k}\cdots\gamma_{i_{N-1}i_{N}}%
^{k}).
\label{coef}
\end{equation}
Here, $\Psi $ is the total wave function of the system,
$\phi $ is the one-electron orbital that is set to the 
Gaussian basis set or the Hubbard site orbital in our calculations,
$\hat{A}$ is the antisymmetrizer,
$N$ is the number of the electron, $M$ is the number 
of the one-electron orbital or of the basis function, and
$K$ is the number of AGPs which appear in the wave function.
The geminal $\gamma $ is a skew-symmetric matrix and there are no other restrictions.
The expression for the total energy with the AGP states of $\gamma ^x$ and $\gamma ^y$ is 
\begin{eqnarray}
E&=&( \sum _{k_1l_1k_2l_2}\frac{1}{2}\mathrm{pf}(1+Bt)\cdot H_{k_1l_1k_2l_2}\nonumber \\ 
&\cdot &((\gamma ^x(1+Bt)^{-1})_{k_2k_1}
(\gamma ^y)_{l_1l_2})t\nonumber \\
&+&\sum _{k_1l_1k_2l_2}
\mathrm{pf}(1+Bt)H_{k_1l_1k_2l_2}\nonumber \\
&\cdot &( -\frac{1}{2}
(B(1+Bt)^{-1})_{l_1k_2}(B(1+Bt)^{-1})_{l_2k_1}\nonumber \\
&+&\frac{1}{2}(B(1+Bt)^{-1})_{l_1k_1}(B(1+Bt)^{-1})_{l_2k_2}\nonumber \\
&-&\frac{1}{2}(B(1+Bt)^{-1}\gamma ^y)_{l_1l_2}(\gamma ^x(1+Bt)^{-1})_{k_2k_1})\cdot t^2)\nonumber \\
&|_{t^{N/2}}& \label{energy}
\end{eqnarray}
where $H$ is the Hamiltonian matrix element and the matrix $B$ is defined as 
\begin{equation}
B = -\gamma ^{y}\gamma ^{x}.
\end{equation}
These are the basic formalisms of extended symmetric
tensor decomposition (ESTD).
These expressions for the AGP states are somewhat similar 
in their structure to expressions found in 
nuclear physics, though they do not contain the 
polynomial expressions\cite{GCM, Onishi1966, doba, Mizusaki2012}.
\par 
Next, we show how numerical stabilization of the energy 
is obtained.
We diagonalize the matrix $B$ as
\begin{equation}
B=Q_r\, M\, Q_l.
\end{equation}
The matrix $B$ is defined as a product of two skew-symmetric matrices and has a relationship with the 
so-called skew Hamiltonian matrices.
Therefore, the eigenvalues of $B$ are evenly degenerate.
For such matrices, the details of the eigenvalue decompositions 
can be found in ref \cite{Ikramov, vanloan1, vanloan2}.
When the eigenvalues of $B$ are $\lambda _1,\lambda _1, \ldots , \lambda _{M/2}, \lambda _{M/2}$, we can rewrite the norm as 
\begin{eqnarray}
n&=&N!\,\mathrm{pf}(1+Bt)|_{t^{N/2}} \nonumber \\
&=&N!\,(\mathrm{det}(1+Bt))^{1/2}|_{t^{N/2}} \nonumber \\
&=&N!\,\sqrt{(1+\lambda _1t)^2\cdots (1+\lambda _{M/2}t)^2}|_{t^{N/2}} \nonumber \\
&=&N!\, ((1+\lambda _1t)\cdots (1+\lambda _{M/2}t))|_{t^{N/2}}.
\end{eqnarray}
We also apply the eigendecomposition of matrix $B$ in the 
energy expression.
This idea was obtained from the description in ref.\cite{coleman2}.
\begin{eqnarray}
E&=&\sum _{k_1l_1k_2l_2}\frac{1}{2}H_{k_1l_1k_2l_2}
\cdot 
((\gamma ^xQ_r\, q1\, Q_l)_{k_2k_1}
(\gamma ^y)_{l_1l_2})\nonumber \\
&+&\sum _{k_1l_1k_2l_2}
H_{k_1l_1k_2l_2}\nonumber \\
&\cdot &(-\frac{1}{2}(Q_rM\times Q_rM\, q2\, Q_l\times Q_l)_{l_1k_2l_2k_1}\nonumber \\
&+&\frac{1}{2}(Q_rM\times Q_rM\, q2\, Q_l\times Q_l)_{l_1k_1l_2k_2}\nonumber \\
&-&\frac{1}{2}(Q_rM\times \gamma ^xQ_r\, q2\, Q_l\gamma ^y\times Q_l)_{l_1l_2k_2k_1})
\end{eqnarray}
where
\begin{equation}
q1_{i_1i_1}=\frac{\partial }{\partial \lambda _{i_1}}\mathrm{pf}(1+Bt)|_{t^{N/2-1}},\label{q1}
\end{equation}

\begin{equation}
q2_{i_1i_3i_1i_3}=\frac{\partial }{\partial \lambda _{i_1}}\frac{\partial }{\partial \lambda _{i_3}}\mathrm{pf}(1+Bt)|_{t^{N/2-2}}.\label{q2}
\end{equation}
Or, alternatively, we have 
\begin{equation}
q10_{i_1i_1}=(1+Mt)^{-1}_{i_1i_1}\mathrm{pf}(1+Bt)|_{t^{N/2-1}},\label{q10}
\end{equation}

\begin{equation}
q20_{i_1i_3i_1i_3}=(1+Mt)^{-1}_{i_1i_1}(1+Mt)^{-1}_{i_3i_3}\mathrm{pf}(1+Bt)|_{t^{N/2-2}}.\label{q20}
\end{equation}
The notations $q10$ and $q20$ are the original formalisms of the energies 
and $q1$ and $q2$ are modified ones which are used instead of $q10$ and $q20$.
For the above expressions, the set of indices $i_1i_1$ or $i_3i_3$
always appear as diagonal and the non-diagonal part is set to zero.
This modification removes the
inverse matrices even though they are given as matrix polynomials.
The tensor given in (\ref{q2}) can be distributed 
as a two-dimensional plane in the four-dimensional space of indices and 
has the twofold structure of the usual diagonal matrices.
The definition of the tensor algebra with the cross term is such that
\begin{eqnarray}
&&(A \times B\,  q\,  C \times D)_{i_1j_1i_2j_2}\nonumber \\
&\equiv &
\sum_{k_1l_1k_2k_2}
A_{i_1k_1}B_{i_2l_1}q_{k_1l_1k_2l_2}C_{k_2j_1}D_{l_2j_2}.
\end{eqnarray}
Equations (\ref{q1}) and (\ref{q10}) give the same value.
However, equations (\ref{q2}) and (\ref{q20}) are different
in the sense that in (\ref{q2}), the $i_1=i_3$ term is neglected.
We numerically found that we could obtain the same value for the 
total energy with this modification in the energy expression.
With this modification, we successfully reproduced
the total energy of STO-3G water converged case.
In this case, the energy was obtained with
quadruple precision arrays in previous research\cite{uemura2015},
but we were able to reproduce almost the same energy
with the same input geminal matrices using double precision arrays.
From this we have concluded that the stabilization of the total energy with double precision
variables is achieved by use of this eigenvalue technique. We can further obtain the algorithm to partially stabilize 
the first-order derivative of the energy with respect to 
geminals in a similar manner.\par
We have applied the ESTD formalism on 
a water molecule and the Hubbard model using the stabilized form of ESTD described above.
We have done the variational process with 
the quasi-Newton  
Broyden-Fletcher-
Goldfarb-Shanno algorithm (BFGS) method.
For the water molecule, we have tested the 
STO-3G basis set and the Dunning zeta (DZ) basis set.
The geometry is set to 
O$:(0.0, 0.0, 0.0)$, H$:(-1.809,0.0,0.0)$, H$:(0.453549,1.75221,0.0)$ for STO-3G
and O$:(0.0, 0.0, -0.009)$, H$:(1.515263,0.0,-1.058898)$, H$:(-1.515263,0.0,-1.058898)$ for DZ.
For the STO-3G case, we have used the same system as that of ref.\cite{uemura2015}. For the DZ basis case, we have used the system in
ref.\cite{saxe}. For the Hubbad model, we have used the 
one-dimensional Hubbard model with six sites. The parameter $U/t$ is set 
to $1.0$ and $10.0$.\par

\begin{table}[tbh]
\begin{center}%
\begin{tabular}
[c]{|l|l|}\hline
Method & Total energy\\\hline
Hartree-Fock & -74.962940033\\\hline
ESTD, $K=1$ & -74.987449763\\
ESTD, $K=4$ & -75.011647636\\
ESTD, $K=8$ & -75.012339655\\
ESTD, $K=16$ & -75.012415900\\\hline
Exact (ours) & -75.012425818\\
Exact (Gaussian) & -75.012425839\\\hline
\end{tabular}
\end{center}
\caption{Total energy (in units of hartrees) of H$_{2}$O with 
STO-3G basis set obtained by ESTD. For
comparison, the full-CI calculation with our own code and the CASSCF
calculation using the Gaussian09 package was taken from ref.\cite{uemura2015}. The Hartree-Fock calculation was done with our own code.}%
\label{tb:h2o_sto}%
\end{table}

\begin{table}[tbh]
\begin{center}%
\begin{tabular}
[c]{|l|l|}\hline
Method & Total energy\\\hline
Hartree-Fock & -76.00983760\\\hline
ESTD, $K=1$ & -76.03854235\\
ESTD, $K=4$ & -76.13814833\\
ESTD, $K=8$ & -76.14419119\\
ESTD, $K=16$ & -76.14774318\\
ESTD, $K=40$ & -76.14971592\\
ESTD, $K=60$ & -76.15509884\\\hline
CISD (ref.\cite{saxe}) & -76.150015\\
CCSD (ref.\cite{purvis}) & -76.156078\\
Exact (ref.\cite{saxe}) & -76.157866\\\hline
\end{tabular}
\end{center}
\caption{Total energy (in units of hartrees) of H$_{2}$O 
with DZ basis set obtained by ESTD. For
comparison, the full-CI calculation was taken from ref.\cite{saxe}.
The Hartree-Fock calculation was done with our own code and 
agrees with the value in ref.\cite{saxe}.
The CISD calculation in ref.\cite{saxe} and the CCSD
calculation in ref.\cite{purvis} are also shown.}%
\label{tb:h2o_dz}%
\end{table}

\begin{table}[tbh]
\begin{center}%
\begin{tabular}
[c]{|l|l|}\hline
Method & Total energy\\\hline
Hartree-Fock & -6.50000000\\\hline
ESTD, $K=1$ & -6.53462699\\
ESTD, $K=4$ & -6.59785021\\
ESTD, $K=8$ & -6.60101260\\
ESTD, $K=16$ & -6.60115439\\\hline
Exact (ours) & -6.60115829\\
Exact (ref.\cite{nakata}) & -6.60115829\\\hline
\end{tabular}
\end{center}
\caption{Total energy of six site Hubbard model with 
U/t=1.0 obtained by ESTD. For
comparison, the full-CI calculation with our own code and the 
result of ref.\cite{nakata} is shown. The Hartree-Fock calculation was done with our own code.}%
\label{tb:hubbard_u1}%
\end{table}

\begin{table}[tbh]
\begin{center}%
\begin{tabular}
[c]{|l|l|}\hline
Method & Total energy\\\hline
Hartree-Fock & -1.18824301\\\hline
ESTD, $K=1$ & -1.26387314\\
ESTD, $K=4$ & -1.51073560\\
ESTD, $K=8$ & -1.65861524\\
ESTD, $K=16$ & -1.66435948\\\hline
Exact (ours) & -1.66436273\\
Exact (ref.\cite{nakata}) & -1.66436273\\\hline
\end{tabular}
\end{center}
\caption{Total energy of six site Hubbard model with 
U/t=10.0 obtained by ESTD. For
comparison, the full-CI calculation with our own code and the 
result of ref.\cite{nakata} is shown. The Hartree-Fock calculation was done with our own code.}%
\label{tb:hubbard_u10}%
\end{table}

Table \ref{tb:h2o_sto} shows our 
results for the water molecule with STO-3G basis set.
The ESTD energy starts between the 
Hartree-Fock (HF) and full-CI values.
Then, the ESTD result rapidly approaches the exact value.
In this system, $M=14$, $N=10$, and the total dimension of the Hilbert space is 
set to $1001$. The residual energy is $1.0 \times 10^{-5}$ hartree when $K=16$. In table \ref{tb:h2o_dz}, we have shown the 
results for the water molecule with the DZ basis set.
In this system, $M=28$, $N=10$, and the total dimension of the Hilbert space is 
$13123110$.
When the spin and space adaptations were done correctly,
the dimension became $256473$\cite{saxe}.
If this was done with the ESTD algorithm, the 
numerical difficulty of searching for the
correct ground state in the total
Hilbert space would be dramatically reduced.
In this system, the variation reaches
the area beyond 95 percent of the total correlation energy.
The ESTD energy of $K=1$ is slightly below the Hartree-Fock energy,
which could be an important sign that the variational space 
for one AGP state is larger than one Slater determinant.
The Configuration Interaction (CISD) and Coupled 
Cluster (CCSD) values are also shown from the references.
Our result of ESTD, $K=60$ is better than the CISD energy. 
The result describes more than 98 percent 
of the correlation energy and closely approaches the CCSD energy.
In this system, the residual energy is $2.8 \times 10^{-3}$ hartree when $K=60$.
The calculation of $K=60$ is done under the condition that each AGP state that appears in eq.(\ref{coef}) has the same weight.
 In tables \ref{tb:hubbard_u1} and \ref{tb:hubbard_u10}, we have shown the 
results for the one-dimensional six-site Hubbard model with $U/t=1.0$
and $U/t=10.0$. 
For both systems, $M=12$, $N=6$, and the total dimension of the Hilbert space is set 
to 924.
In both cases, the ESTD energy is well-converging to the exact value.
For the system with $U=1.0$, the residual energy is $3.9 \times 10^{-6}$ when $K=16$.
For the system with $U=10.0$, the residual energy is $3.3 \times 10^{-6}$ when $K=16$.
\par

\begin{figure}
[ptb]
\begin{center}
\includegraphics[height=3.0in, width=3.0in]
{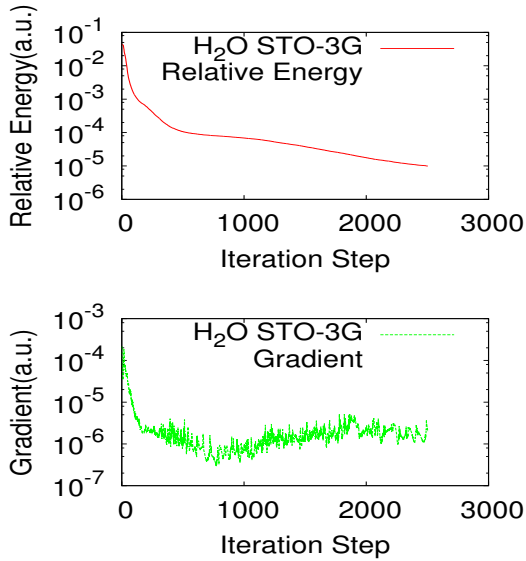}
\caption{The relative energy and the energy gradient on the variation of 
the water molecule with STO-3G basis set. The number of AGPs ($K$) is set to 16.}
\label{figm14grad}
\end{center}
\end{figure}

\begin{figure}
[ptb]
\begin{center}
\includegraphics[height=3.0in, width=3.0in]
{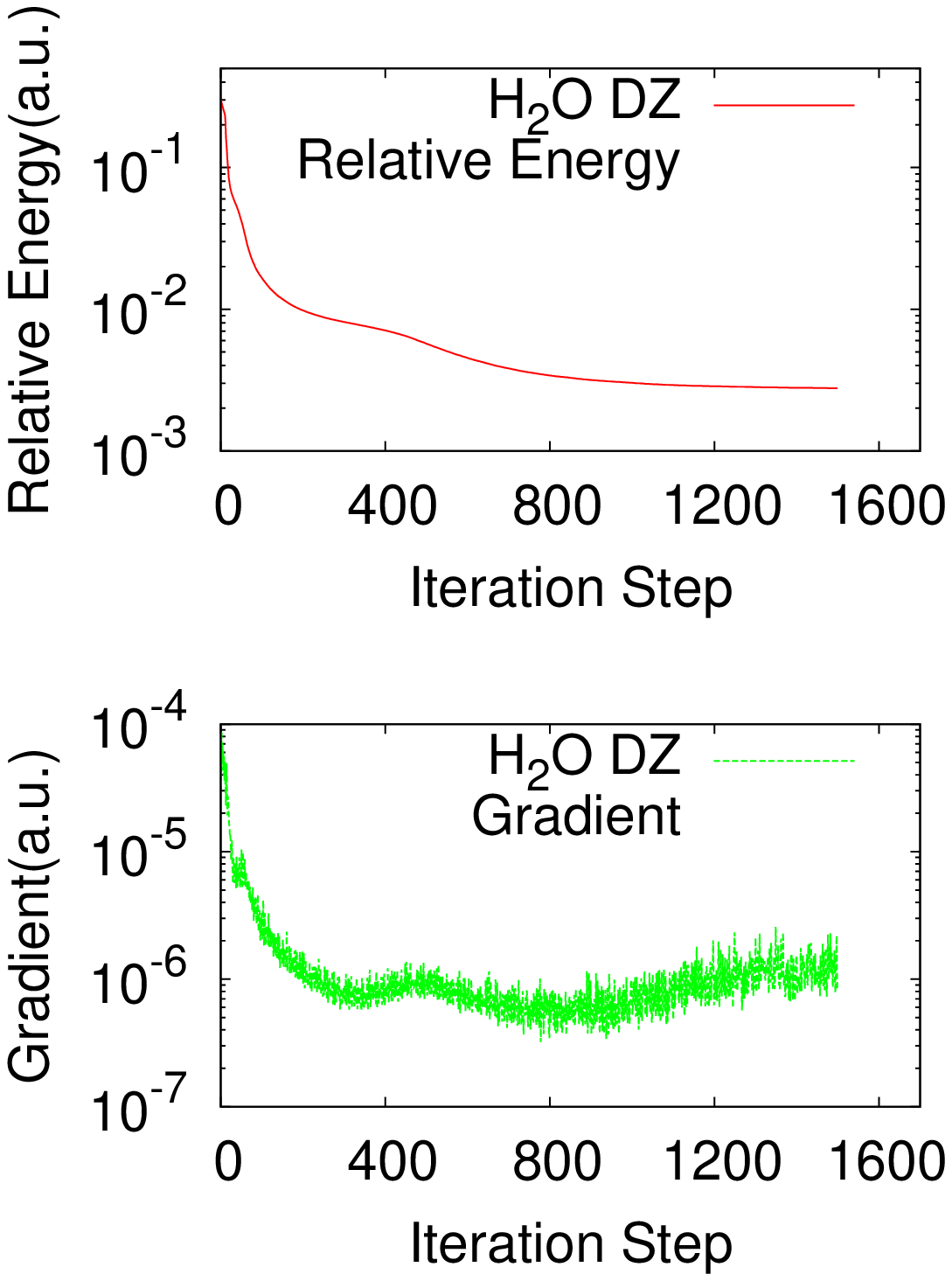}
\caption{The relative energy and the energy gradient on the variation of 
the water molecule with DZ basis set. The number of AGPs ($K$) is set to 60.}
\label{figm28grad}
\end{center}
\end{figure}

\begin{figure}
[ptb]
\begin{center}
\includegraphics[height=3.0in, width=3.0in]
{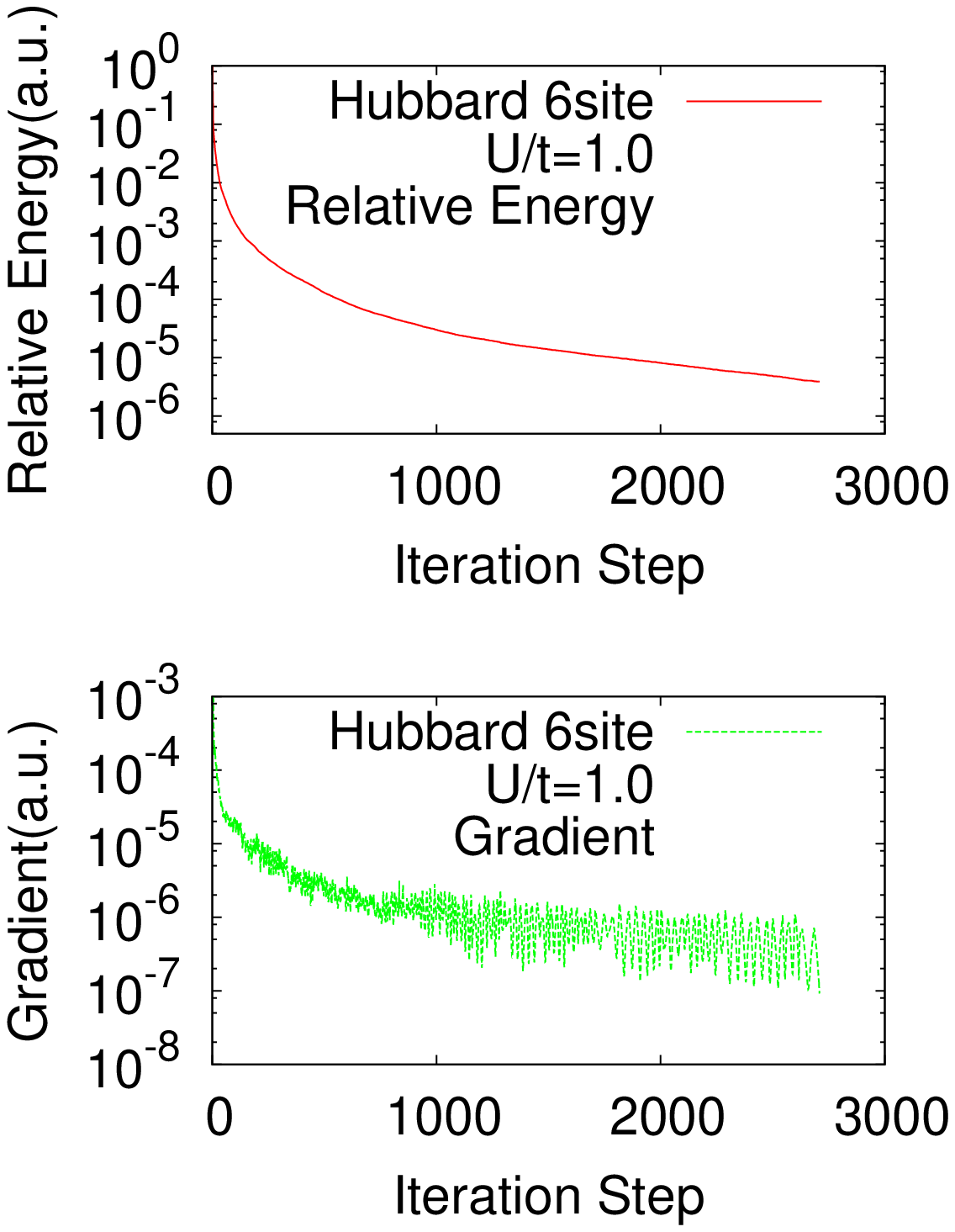}
\caption{The relative energy and the energy gradient on the variation of 
the one dimensional six site Hubbard model with $U/t = 1.0$. The number of AGPs ($K$) is set to 16.}
\label{figm12u1grad}
\end{center}
\end{figure}

\begin{figure}
[ptb]
\begin{center}
\includegraphics[height=3.0in, width=3.0in]
{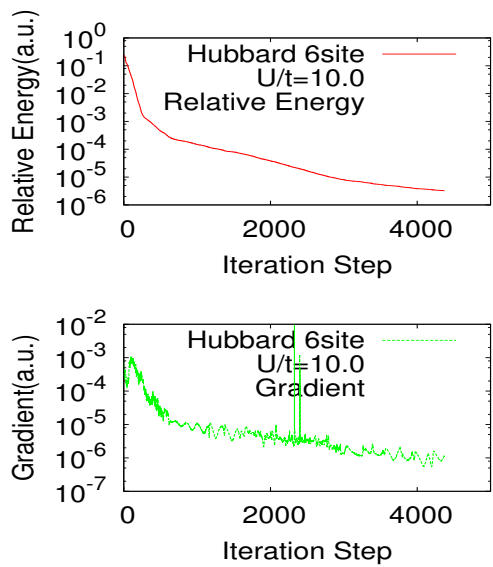}
\caption{The relative energy and the energy gradient on the variation of 
the one dimensional six site Hubbard model with $U/t = 10.0$. The number of AGPs ($K$) is set to 16.}
\label{figm12u10grad}
\end{center}
\end{figure}

In Fig.\ref{figm14grad}-\ref{figm12u10grad}, we have
plotted the behavior of relative energy with
respect to exact value and the energy gradient 
with respect to the geminal variables.
We have used the root mean square for the value of the geminal gradient.
In each case, the value of the gradient is also decreasing 
as the variation progresses.
These first-order derivatives are combined with 
the BFGS Hessian matrices and determine the 
direction of the variational search of the geminals.\par

We have observed that by the application of the 
eigenvalue technique with AGPs, the instability of the energy and
some of the variational process is removed.
We were therefore able to perform the ESTD calculations
with double precision arrays and general skew-symmetric 
matrices without the usage
of quadruple or higher arithmetic.
As a result, the ESTD calculation of 
general matrices became dramatically faster than before.
The energy error for the DZ basis case is 
around three milihartree in this case, which is comparable to the error in the Hilbert space JAGP case\cite{eric2}.
This energy error is likely to be further decreased when
we increase the number of AGPs.
For the case of the Hubbard models,
the behavior of residual energies shows
that the exponential convergence toward the exact solution is likely
against the number of terms in ESTD.
We have not optimized our formalism for spin adaptations,
but if that is done correctly, we expect overall improvement
of the variation.
Also, we could calculate the excited states 
within the ESTD formalism by using the 
orthogonality conditions of the quantum states.
All calculations apart from the $K=1$ case have 
been done with the parallelization for the $K^2$ part of
the ESTD algorithm.\par

We acknowledge the Strategic Programs for Innovative Research by the Ministry
of Education, Culture, Sports, Science and Technology of Japan and “Priority Issue on Post-K Computer” (Development of new
fundamental technologies for high-efficiency energy creation, conversion/storage and use) for financial support during
our research. Some of the computations in the present study were performed using the Research Center for Computational Science, Okazaki, Japan.

\end{document}